# Vertical CNT-Si Photodiode Array

*Arman Ahnood [1,2], Hang Zhou [3*], Qing Dai [1], Yuri Vygranenko[4], Yuji Suzuki [5], MR Esmaeili-Rad [6], Gehan Amaratunga [1], Arokia Nathan [1**]*

[1] Electrical Engineering Division, Department of Engineering, University of Cambridge, Cambridge, United Kingdom

[2] School of Physics, University of Melbourne, Melbourne 3010, Australia

[3] Peking University Shenzhen Graduate School, Peking University, Shenzhen, China

[4] Electronics Telecommunications and Computer Engineering, Instituto Superior de Engenharia de Lisboa, Lisboa, Portugal

[5] London Centre for Nanotechnology, University College London, London, United Kingdom

[6] Department of Electrical Engineering and Computer Sciences, University of California, Berkeley, California, United States

Corresponding authors: * Hang Zhou: zsuchatman@gmail.com , ** Arokia Nathan: an299@cam.ac.uk

*A photodiode consisting nanopillars of thin film silicon p-i-n on an array of vertically aligned carbon nanotubes (CNT) with a non-continuous cathode electrode is demonstrated. The structure exploits the intrinsic enhancement of the CNTs' electric field, which leads to reduction in the photodiode's operating voltage and response time, and enhancement of optical coupling due to better light trapping, as compared with the conventional planar photodiode. These improvements translate to higher resolution and higher frame rate flat panel imaging systems for a broad range of applications, including computed tomography and particle detection.*

High scalability, low cost, and resilience to high-energy particles are some of the attributes of thin film silicon photodiode, which have led to its use in a diverse range of imaging applications. In particular, hydrogenated amorphous silicon (a-Si:H) flat panel imagers (FPI), which consist of an array of photodiodes and associated amplification and readout circuitry, have found usage as x-ray detectors in computed tomography (CT)[1]. Here, due to the difficulties in x-ray focusing[2], a detector size similar to the subject is desirable. Detection of high-energy beta particles[3] or neutrons[4] is another application of FPIs, where the radiation resistance of a-Si:H[5] extends detector lifetime.

Despite these benefits, a-Si:H based photodiodes have two performance limiting drawbacks from a fundamental materials prospective. The disordered nature of the a-Si:H photo-absorber manifests itself as deep-states within its energy gap[6] acting as trapping centers for photo-generated carriers. Thermal re-emission[7] of trapped carriers in the absence of illumination gives rise to a persistent photocurrent, with a decay time scale of ~100 secs[7,8], thus limiting the photodiode's response speed to illumination changes. This leads to image lag effects[8]. The disorder also gives rise to photocarrier recombination losses. While this can be mitigated though use of thinner photoabsorber layers, concurrently favouring high collection efficiency of photo-generated carriers, an issue arises at high wavelengths in which the reduced thickness becomes less than the penetration depth yielding incomplete light absorption. In contrast, a larger absorber thickness requires a high reverse bias voltage across the photodiode. Despite reports of carrier collection efficiencies in excess of unity[9–11] at high voltages, this approach is undesirable due to possible device breakdown[10,11]. Furthermore this would require design of high voltage circuits using conventional low cost and large area thin film electronics for heterogeneously integration with photodiode array[12].

Apart from the aforementioned fundamental materials limitations, there remain several device design challenges associated with the use of a-Si:H photodiodes, and its implications on the resolution of the FPI. The photodiode's photocurrent needs to be higher than the electrical noise of the FPI at system level. Despite the development of various readout circuits with amplification functionality[13] this constraints the minimum photodiode area, pixel size, and therefore the maximum FPI resolution, particularly for low light level detection applications.

Vertical CNT arrays have already been used in conjunction with a-Si:H solar cells to enhance their efficiency[14,15] through enhanced optical absorption especially at higher wavelengths[15] due to light trapping[16] and vertical absorption path[17]. The vertical CNT structure with radial p-i-n diode enables the photons to be absorbed in the vertical direction while photogenerated carriers are separated and collected in the radial direction. This effect has also been exploited in core-shell type nanowire photovoltaic device [18,19]. Moreover, the optics of CNT arrays could be optimized for broadband light absorption by adjusting the CNT pitch and the length of the nanotube arrays[20]. CNTs have also been used as a means of electric field enhancement within the solar cell[21–23]. This letter presents an array of vertical cylindrical a-Si:H p-i-n photodiode nanopillars using an array of vertical CNTs for enhanced light trapping and electric field with implications for improved spatial resolution and speed of image capture.

Fig. 1 a) shows the cross-sectional schematic of the device fabricated in this work. The cylindrical photodiodes were formed by consecutive deposition of p-, i- and n-type a-Si:H layers over a 2×2 mm array of vertically aligned multi-wall CNTs on a c-Si substrate (see Fig. 1(b)). Prior to the growth of CNTs, a thermal silicon oxide layer was grown on the c-Si substrate, followed by deposition of a tungsten film. The surface of the tungsten film was subsequently oxidized prior to deposition of the a-Si:H layers, forming a semi-insulating film which acts as a barrier against transport of charge carriers in the planar regions of the device. This approach ensured that the planar regions of the device have negligible contribution to the overall current-voltage characteristics, whilst allowing the vertical photodiodes to be probed through the conductive CNT pillars interconnected via the conductive tungsten film below the thin tungsten oxide. A top semi-transparent titanium/gold film was evaporated through a shadow mask with 3×3 mm openings to define the photosensitive area of the CNT-Si photodiode array. It should be noted that the evaporation, when compared with sputtering, takes place at high vacuum, leading to highly directional deposition with minimal sidewall coverage. This allows creation of CNT photodiode sidewalls with minimal metallic Ti/Au layer coverage, as depicted in Fig. 1 c). For comparison, a reference planar p-i-n structure was also fabricated on an area of the substrate where CNTs had intentionally not been grown. Further information on the fabrication processes is presented in the supplementary section.

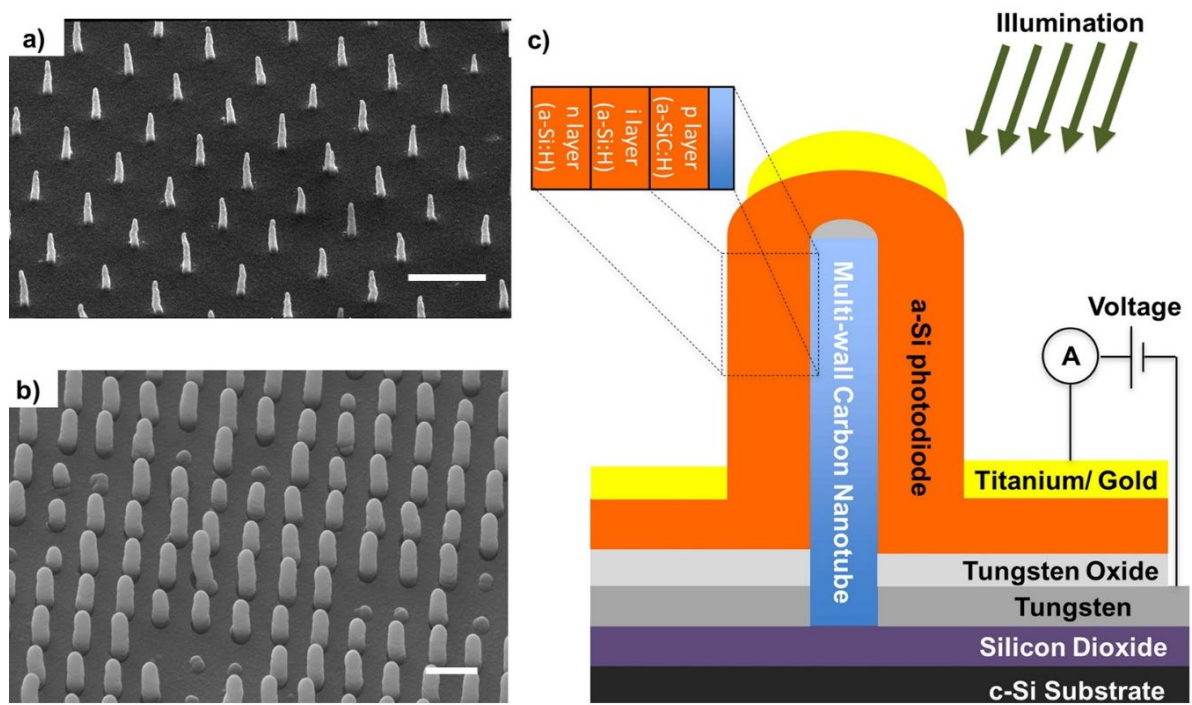

**Figure 1. a)** SEM image of the as grown array of vertical CNTs with 1μm height, 80nm diameter and 800nm separation and **b)** CNTs arrays covered with thin film silicon p-i-n and Ti/Au layers. **c)** Schematic diagram of the cross sectional view of a single photodiode pillar and the associated measurement arrangement.

The dark current of the CNT-Si photodiode array is governed by the sum of two components: contribution from the vertical CNT photodiode nanopillars and contribution from the planar region. As outlined earlier, the fabrication steps undertaken here minimize the contribution of the planar region, thus allowing the probing of the vertical CNT pillar diodes. The effectiveness the semi-insulating tungsten oxide layer as a means of achieving this can be examined by considering Figure 2 a), which compares the dark current-voltage characteristics of the planner device with the CNT-Si device. As shown in Figure 2 a), the planar device's dark current in forward biases is reduced by over two order of magnitude when compared to that of the CNT-Si photodiode array, which indicates that the dark current in the CNT-Si photodiode array device is dominated by the current though the vertical CNT-Si photodiode nanopillars. It should be noted that that at high voltages, the dark current voltage characteristics is dominated by the large series resistance which influences the ideality factor of both the photodiodes. Applying the conventional method for ideality factor extraction to Figure 2 a) yields ideality factor of 14.5 for the CNT-Si photodiode array compared with 22 for the planar device, indicating that the presence of semi-insulating tungsten oxide leads to an increase in the series resistance of the planar regions of the CNT-Si photodiode array. This points to the effectiveness of the semi-insulating tungsten oxide layer as a means of suppressing the contribution of the planar region of the CNT-Si photodiode array to the current-voltage characteristics. It should be noted that the p-i-n photodiode structure selected for this work leads to devices characteristics that are independent of work functions of CNT and Ti/Au contacts[24]. This is because sufficiently high dopant density of the n- and p-layers creates a metal-semiconductor barrier width which is sufficiently narrow for the junction to exhibit ohmic behavior.

Although both devices occupied the same lateral area, the active p-i-n junction area of the planar and CNT-Si photodiode array are not equal. However the simple increase in the junction area by the virtue of the additional surface created by the CNT pillars is insufficient to explain the two order of magnitude increase in dark current. The array used in this work has about $6.25 \times 10^6$ individual CNT elements with 1µm height and 40nm radiance, resulting in a 41% increase in the anode's surface area, as supported by the area ratio extracted from the photodiode junction capacitance measurements (see supplementary information). Cracks and microvoids across the photodiode provide shunt paths across the device, and could be an alternative mechanism for the increase in the dark current. Although this explanation could potentially support the observed increase in the dark current, the increase in open circuit voltage (i.e. voltage across the device when current output is minima) shown in Figure 2(b)) is inconsistent with this mechanism. The open circuit voltage in the CNT-Si photodiode array is found to be 0.6V, compared to 0.5V observed in the planar device. The higher open circuit voltage observed in the vertical CNT-Si photodiode array compared to that of the planar structure can be attributed to the larger carrier collection in the vertical CNT photodiode nanopillars. Further discussion on this will be presented later in this letter. Measurements were performed using a halogen lamp with illumination intensity of 1.2mW/cm$^2$, yielding diode efficiency of 7.5% at zero bias.

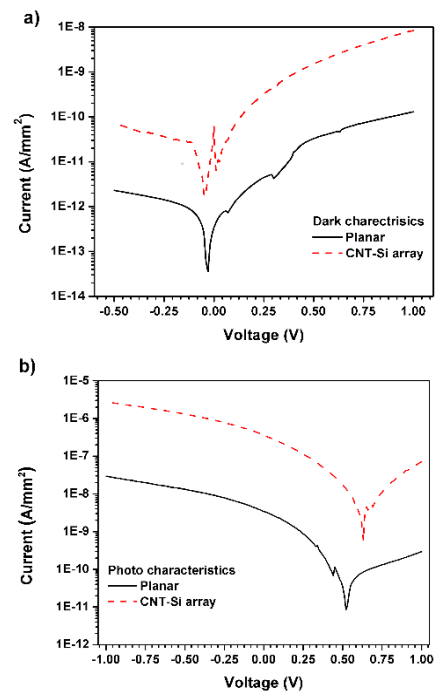

**Figure 2. a)** and **b)** the current voltage characteristics of the planar and CNT-Si photodiode array structures measured in dark and under illumination, respectively. The planar device has significantly lower current compared with the CNT-Si photodiode array, indicating role of semi-insulating tungsten oxide in effectively suppressing the contribution of the planar regions of the CNT-Si photodiode array to the current-voltage characteristics.

Up to this point we have demonstrated that the contribution of the planar regime of the CNT-Si photodiode array device to the current-voltage characteristics of the device is negligible compared to the CNT-Si photodiode nanopillar region. Thus we can now investigate the properties of the vertical CNT-Si photodiode array discounting the contribution of the planar areas. The fabrication process adopted in this work produces vertical photodiodes without the sidewall metallization. The rest of this paper investigates the consequence of the metal free sidewalls.

It is well established that the convergence of electric field lines at the convex surface of the metallic CNT pillar leads to a localized enhancement of the electric field[25], which serves to increase device conductivity[26]. Although the CNT arrays in the photovoltaic devices have been widely used in earlier works, the reports of electric field enhancement effect have been limited to certain device configurations[22]. The origin of this inconsistency can be explained by considering the sputter deposited transparent conducting oxide (TCO) to achieve a continuous top electrode[14,15] used in conventional approaches. The presence of a concave internal surface at the photodiode pillar's sidewalls leads to quenching of the CNT's electric field enhancement. In this work this is resolved through the use of evaporated Ti/Au top metal electrodes, which eliminates metallization of the concave internal surface at the surface of the photodiode pillar's sidewalls (see Figure 1 a)). Figure 3 compares the simulated electric field distribution of a CNT pillar photodiode with metallic sidewall shown in Figure 3 a), with that without as shown in Figure 3 b). The cross-sectional schematic shows that the metallic sidewall leads to a reduction in the electric field near the n-layer from $10^4$ to $10^3$ V/cm. This is further discussed in the supplementary section.

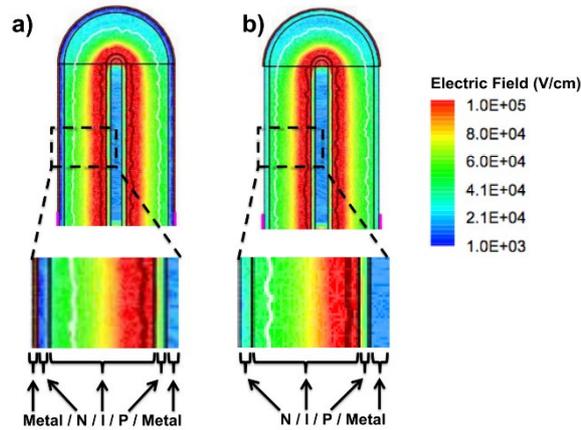

**Figure 3.** Cross sectional diagram of a 3-D electric field profile simulated for two CNT photodiodes: one with sidewall cathode **a)** and second without sidewall cathode **b)**. Although the use of CNT leads to electric field enhancement in the i-layer for both structures, use of sidewall electrode leads to over one order of magnitude reduction in the electric field in the n-layer of CNT photodiode with sidewall cathode compared with that without sidewall cathode (i.e. from $10^4$ to $10^3$ V/cm).

Further device characterization includes current-voltage measurements between the tip of a single CNT photodiode pillar and back electrode using conductive-tip AFM shown in Figure 4. The equivalent circuit of the device can be presented as a network of discrete vertical CNT photodiode pillars as shown in Figure 4(a), where the resistors in the circuit stem from the non-continuous Ti/Au electrode. The results presented so far have been based on current-voltage measurements performed between the planar Ti/Au and tungsten electrodes, i.e. "configuration A" in Figure 4(a), where the series resistance limits the current thus influencing the current voltage characteristics of the device. Figure 4(b) shows the power law dependence of photocurrent on voltage when the measurement is performed between the tip of a single CNT photodiode nanopillar and back tungsten electrode, i.e. "configuration B" in Figure 4(a) using c-AFM. Although the illumination intensity at the CNT photodiode nanopillar is expected to be distorted by the shadowing of the c-AFM tip directly above the photodiode and therefore cannot be determined, nevertheless a comparison can be made between the power law dependence of photocurrent on voltage in contrast with that in "configuration A" where the variation in reverse-biased photocurrent with voltage is less pronounced as shown in Figure 2 (b). This power law dependence is consistent with the power law expected in the conductivity of a-Si:H at high electric fields. At high fields, the dependence of a-Si:H conductivity ($\sigma$) on electric field (F) follows a super-linear relation $\sigma=F^{\gamma}$, where $\gamma$ is an empirical parameter in the range of 2 to 3[27].

The $V_{OC}$ measured by the c-AFM on a single nanopillar is found to be ~0.75V, which exceeds the $V_{OC}$ ~0.6V in the whole CNT-Si photodiode array device and ~0.5V for the planar device. Both of these results are consistent with the simulation result presented in Figure 3, which predicts a higher electric field in the absence of metallic sidewall leading to a higher $V_{OC}$ in the single nanopillar diode as well as in the power law dependence of its current-voltage characteristics. It is a fact that the whole CNT-Si photodiode array is composed of parallelly-connected nanopillar diodes and planar diodes. The $V_{OC}$ of the whole device therefore falls between the $V_{OC}$ of the planar diode (~0.5V) and that of the nanopillar diode (~0.75V).

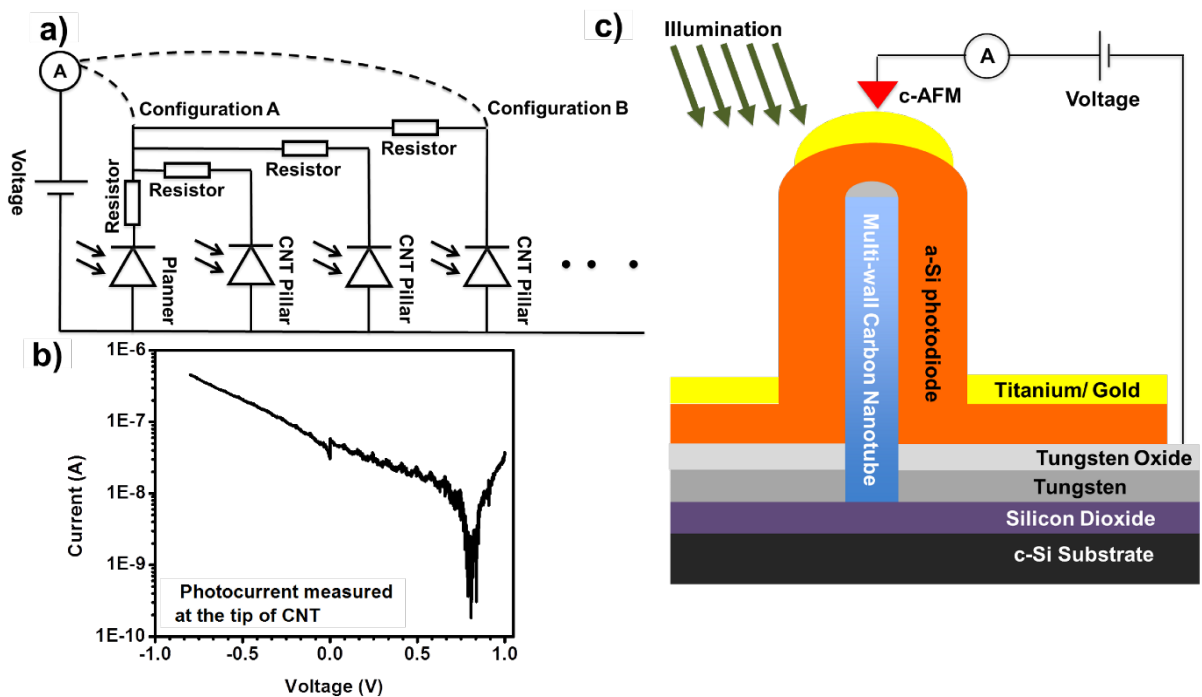

**Figure 4. a)** Equivalent circuit of the CNT array structure consisting of an array of parallelly-connected CNT pillar photodiode segments. It should be noted that the planar photodiode segment is neglected due to its negligible contribution. The resistors represent the access resistance across the n-type a-Si:H films covering the side wall of the CNT pillar photodiodes. This resistance is significant in the absence of sidewall metallization. In the measurement configuration A, the contact is made between the tungsten bottom contact and the planar section of the CNT array device. In configuration B the measurement is performed between the tungsten bottom contact and the tip of the CNT pillar photodiode segment using c-AFM. **b)** Current-voltage characteristics of the c-AFM measurement performed under illumination. The power law variation in the photocurrent with voltage in the reverse bias regime is consistent with the conductivity model of a-Si:H under high electric fields. **c)** Cross sectional diagram of the c-AFM measurement performed between the nickel bottom contact and the tip of the CNT pillar photodiode segment.

Figure 5 shows the transient current response of the CNT-Si array to pulsed illumination. The CNT-Si photodiode array device shows a significantly short decay time of less than 5 secs, compared to the 100s of sec reported elsewhere[7]. This short decay time can be attributed to the increase in the re-emission rate of trapped photo-generated carriers under high electric field[28,29]. The temperature/electric field analogy[30,28] for charge trapping in the a-Si:H implies that the decay time values measured at elevated temperatures in a-Si:H photodiodes[31] are in agreement with our high-electric field results at room temperature.

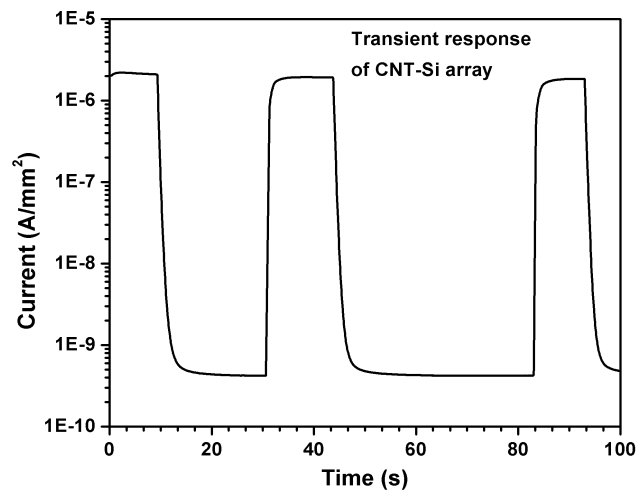

**Figure 5.** Transient response of the CNT-Si photodiode array subject to pulsed illumination. The CNT-Si photodiode array exhibits photocurrent decay rate of the order of 5 secs, significantly faster than conventional a-Si:H photodiodes (see supplementary Figure S6).

This work demonstrates an a-Si:H p-i-n photodiode fabricated on an array of vertical CNTs. The electric field enhancement effect associated with use of vertical CNT-Si pillar leads to higher dark- and photo-current densities at low voltage, and a faster photodiode response time. Based on these results, orders of magnitude improvement in resolution and speed can be realized (this is further discussed in the supplementary information). Large area, high-speed, high-resolution detection technology has wide ranging implications for low-dosage computed tomography scan or enhanced particle detection.

Acknowledgement

Hang Zhou would like to acknowledge National Natural Science Foundation of China (61204077) and Shenzhen Science and Technology Innovation Commission (JCYJ20120614150521967).